\documentstyle[prb,aps,twocolumn]{revtex}
\input epsf.tex
\begin{document}
\draft

\title{Electron-phonon coupling and Peierls transition in metallic carbon nanotubes}
\author{A. S\'{e}d\'{e}ki, L. G. Caron, and C. Bourbonnais}
\address{Centre de Recherche en Physique du Solide, D\'{e}partement de Physique,\\
Universit\'{e} de Sherbrooke, Sherbrooke, Qu\'{e}bec, Canada J1K 2R1}
\date{\today}
\maketitle

\begin{abstract}
We reexamine the putative Peierls transition in a
(5,5) metallic nanotube. We show that the conduction electrons at the Fermi
level do not couple to the longitudinal acoustic phonon but rather to a
folded-in graphene zone edge phonon having the proper K\'{e}kul\'{e}
modulation symmetry. The calculation for the mean-field transition
temperature gives 15 K, a value comparable to previous estimates. We discuss
the significance of this transition temperature.
\end{abstract}


\preprint{HEP/123-qed}

\narrowtext

Since their discovery by Ijima\cite{ijima} in 1991, carbon nanotubes have
attracted a lot of interest due to their unusual geometry and their
structural and electronic properties. Band calculations \cite{theorie}
(confirmed by experiments\cite{Dunlap,exp}) have stressed the
one-dimensional (1D) character of single-wall carbon nanotubes (SWCNT). They
showed moreover that, depending on their diameter and heliticity, a SWCNT
can either be metallic or semi-conducting. The metallic tubes provide an
excellent environment for one-dimensional (1D) conductor physics. The
existence of a putative Peierls transition due to the electron-phonon
coupling has already been addressed by Mintmire \cite{Dunlap} and\ Huang 
\cite{Solide} who predicted the occurrence of a Peierls transition at 1K and
9K respectively. We propose to reexamine the Peierls transition owing to
what we believe to be a wrong selection of soft phonon mode by Huang. The
longitudinal polarization of the phonon they considered, taken from \cite
{PRB48}, cannot lead to a K\'{e}kul\'{e} instability natural to graphene,
the two-dimensional (2D) carbon sheets that are rolled up to form the
tubules. We will consider in the following the (5,5) armchair SWCNT which
presents metallic properties. We shall reevaluate its phonon spectrum,
deduce the electron-phonon coupling amplitude, and finally estimate the
mean-field Peierls transition temperature.

We first recalculate the phonon spectrum of graphene by the dynamical matrix
method using the force constants obtained by Maeda\cite{maeda}. We solve
numerically the secular equation for the phonons. The polarization vector of
the phonons is written as

\begin{equation}
{\bf e}\left( \overrightarrow{q},\nu \right) =\left( {\bf e}_{A}\left( 
\overrightarrow{q},\nu \right) ,{\bf e}_{B}\left( \overrightarrow{q},\nu
\right) \right) \text{, } 
\end{equation}
where ${\bf e}_{r}\left( \overrightarrow{q},\nu \right) $ is the
polarization of one of the two atoms $r=A,B$ in the unit cell for the phonon
of wave vector $\vec{q}$ and branch $\nu $. Here ${\bf e}_{r}$ has two
components along the mutually orthogonal directions $\hat{x}$ and $\hat{y}$
(the tube axis direction) in the sheet respectively (see Fig. 1). These
polarizations satisfy the relation ${\bf e}\left( -\overrightarrow{q},\nu
\right) ={\bf e}\left( \overrightarrow{q},\nu \right) ^{\ast }$.

\epsfxsize=7cm
\begin{figure}
\centerline{\epsfbox{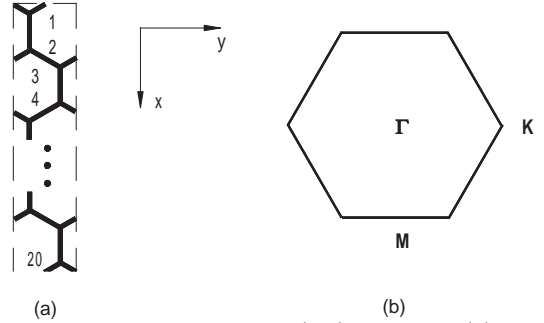}}

\caption{ Unfolded unit cell of (5,5) nanotube (a) and Brillouin
zone of graphene (b).}
\label{}
\end{figure}

There are two phonon groups relevant to our study. The first one is at the $%
K $ points in the 2D hexagonal 
first Brillouin zone (BZ) of graphene (see Fig. 1). Its phonons have the
wave vector $\overrightarrow{q}=\left( 0,\pm 4\pi /3a\right) $ or $\left(
\pm 2\pi /\sqrt{3}a,\pm 2\pi /3a\right) $. The polarizations for $%
\overrightarrow{q}=\left( 0,4\pi /3a\right) $ are given by
\begin{eqnarray}
{\bf e}\left( \overrightarrow{q},1\right) = && \frac{1}{2}\left( i,1,-i,1\right) 
\text{, }{\bf e}\left( \overrightarrow{q},2\right) =\frac{1}{2}\left(
-i,1,i,1\right),\cr
{\bf e}\left( \overrightarrow{q},3\right) = && \frac{1}{2}\left(
-i,-1,-i,1\right) \text{, }{\bf e}\left( \overrightarrow{q},4\right) =\frac{1%
}{2}\left( i,-1,i,1\right)
\end{eqnarray}
The second group, to which the acoustic phonon with longitudinal
polarization belongs, has polarizations
\begin{eqnarray}
{\bf e}\left( \overrightarrow{q},1\right) = &&\frac{1}{\sqrt{2}}\left(
0,1,0,-1\right),
{\bf e}\left( \overrightarrow{q},2\right) =\frac{1}{%
\sqrt{2}}\left( -i,0,i,0\right) \cr
{\bf e}\left( \overrightarrow{q},3\right) = && \frac{1}{\sqrt{2}}\left(
0,1,0,1\right) \text{, }{\bf e}\left( \overrightarrow{q},4\right) =\frac{1}{%
\sqrt{2}}\left( i,0,i,0\right) 
\end{eqnarray}
for $\overrightarrow{q}=\left( 0,-2\pi /3a\right) $. Here $a/\sqrt{3}$ is
the carbon-carbon distance in graphene.

The nanotube is constructed by rolling the sheet of graphene in the
prescribed fashion\cite{theorie}. The (5,5) SWCNT has the right diameter to
be capped at the ends by half a $C_{60}$ sphere. Its unit cell has length $a$
and contains 20 atoms (see Fig. 1). The requirement of periodicity around
the nanotube imposes the following quantization condition on $q_{x}=\left(
2\pi /5\sqrt{3}a\right) m$, where $-5<m\leq 5$ is an integer. The
equilibrium position of an atom on the nanotube is specified by $\left(
n,s\right) $, $n$ indicating the position $R_{n}$ of the unit cell along the
nanotube axis and $s$ referring to the position $\stackrel{\rightarrow }{%
d_{s}}$ of the atom within the unit cell, so that

\begin{equation}
\vec{R}_{n,s}=R_{n}\hat{y}+\vec{d}_{s}. 
\end{equation}
Because of the difference in the length of the unit cell of SWCNT (5,5) with
the one of graphene, the first Brillouin zone has a smaller length in the $%
\hat{y}$ direction given by $2\pi /a$. We can than approximate the phonon
spectrum of the nanotube by folding the spectrum of graphene, subject to the
quantization rule for $q_{x}$ above, into the BZ of the nanotube. We can now
write the polarization vector ${\bf S}_{s}\left( q,\nu ,m\right) $ of atom $%
s $ for the phonon of wave number $q$ (along the axis of the tube)
corresponding to branch $\nu $ and quantum number $m$ in terms of those in
graphene:
\begin{eqnarray}
{\bf S}_{1(2)}\left( q,\nu ,m\right) = && \frac{1}{\sqrt{10}}{\bf e}%
_{A(B)}\left( \overrightarrow{q},\nu \right) \text{,} \cr
 {\bf S}_{3(4)}\left( q,\nu ,m\right) = &&\frac{e^{i\left( q+G_{q}\right) \frac{a%
}{2}+iq_{x}\frac{a\sqrt{3}}{2}}}{\sqrt{10}}{\bf e}_{A(B)}\left( 
\overrightarrow{q},\nu \right) \text{,} 
\end{eqnarray}
where $q_{y}=q+G_{q}$ $\ $for $\left| q\right| \leq \pi /a$, and $%
G_{q}=n_{q}(2\pi /a)$ is a reciprocal lattice wave number of the nanotube.
Also, one has the property ${\bf S}_{s+4}=\exp (iq_{x}\sqrt{3}a){\bf \ S}%
_{s} $. The result of this folding-in for the branches of interest here is
shown in Fig. 2.

The electron-phonon interaction Hamiltonian, treated within the
tight-binding approximation, is written as\cite{Solide}

\begin{eqnarray}
H_{e-ph} &&=\frac{1}{\sqrt{L}}\sum_{v,m,p,p^{\prime },k,k^{\prime },\sigma
}g_{k,p;k^{\prime },p^{\prime }}^{v,m}(G)C_{k,p,\sigma }^{+}C_{k^{\prime
},p^{\prime },\sigma }\cr
&&\times \left( b_{q,\nu ,m}+b_{-q,\nu ,m}^{+}\right) \text{,}
\end{eqnarray}
where $L$ is the length of the tubule, $C_{k,p,\sigma }^{+}$ ($C_{k,p,\sigma
}$) is the creation (annihilation) operator for an electron with wave vector 
$k$ and spin $\sigma $ in band $p$, $G\ $is a reciprocal lattice wave number
such that $q=k-k^{\prime }-G$ is within the natotube's first BZ, and $%
b_{q,\nu ,m}$ is the phonon operator. The electron-phonon interaction matrix
element $g_{k,p;k^{\prime },p^{\prime }}^{v,m}(G)$, is given within the
tight-binding approximation by the following expression : \vspace{-0.2cm}
\begin{eqnarray}
g_{k,p;k^{\prime },p^{\prime }}^{v,m}(G) && =-\left( \frac{\hbar \omega
_{k-k^{\prime }-G,\nu ,m}}{2\mu }\right) ^{\frac{1}{2}}\cr
 && \sum_{\alpha ,n\prime
,s,s\prime }[\nabla _{\alpha }J(0,s;n^{\prime },s^{\prime
})]_{0}U_{s,p}^{\ast }(k)U_{s^{\prime },p^{\prime }}(k^{^{\prime }}) \cr
&& \times \ \{e^{-ik^{\prime }R_{n\prime }}\left[ {\bf S}_{s}\left( k-k^{\prime
}-G,\nu ,m\right) \right] _{\alpha } \cr
&& - \ e^{-ikR_{n^{\prime }}}\left[ {\bf S}%
_  {s^{\prime }}\left( k-k^{\prime }-G,\nu ,m\right) \right] _{\alpha }\}\text{%
,}
\end{eqnarray}
 where $\mu $ is the mass per unit length of the tubule, $\omega
_{q,\nu ,m}$ is the phonon frequency, $J(n,s;n^{^{\prime }},s^{^{\prime }})$
is the matrix element of the atomic potential connecting orbitals on
nearest-neighbor atoms $(n,s)$ and $(n^{\prime },s^{\prime })$, and $U(k)$
is the Bloch eigenvector that diagonalizes the purely electronic part of
Hamiltonian. The elements of $U(k)$ obey to the periodic boundary condition $%
U_{1,p}(k)=U_{21,p}(k)$ and correspond to the graphene Bloch eigenvectors
folded into the nanotube's BZ. The Fermi level is situated at $k_{F}=\left(
2\pi /3a\right) $ at the crossing of two bands having the dispersion
relations $\varepsilon _{p}(k)=\pm J_{0}\left[ 1-2\cos (ka/2)\right] $ with $%
p=1,2$. The band crossing corresponds to the $K$ point of the BZ of graphene
of Fig. 1b. The wave functions at the six $K$ points bear the signature of
the K\'{e}kul\'{e} bond arrangement of graphene which can be obtained by
coupling to the $\overrightarrow{q}=\left( 0,\pm 4\pi /3a\right) $, $\left(
\pm 2\pi /\sqrt{3}a,\pm 2\pi /3a\right) $ phonons. It is now possible to
calculate the electron-phonon coupling amplitudes for these two conduction
bands by using the eigenvectors $U(k)$ given in the article of Jishi \cite
{PRB48}.

It is important to realize at this point that the momentum conservation
condition $q=k-k^{\prime }-G$ imposes that $G\neq 0$ whenever $\left|
k\right| =\left| k^{\prime }\right| =k_{F}$ since $\left| k-k^{\prime
}\right| =4\pi /3a$ is outside the first BZ of the nanotube and must be
brought back within by a $\left| G\right| =2\pi /a.$ As a consequence of
this, one has $\left| q\right| =2\pi /3a$. There are two possibilities of
interest here. The first corresponds to a graphene phonon having $\left|
q_{y}\right| =\left| q+G_{q}\right| =4\pi /3a$ for $m=0$ or $\left|
q_{y}\right| =\left| q\right| =2\pi /3a$ for $m=5$. These are the $K$ point
phonons of graphene. The other is $\left| q_{y}\right| =\left| q\right|
=2\pi /3a$ for $m=0$ to which the acoustic phonon with longitudinal
polarization used by Huang\cite{Solide} and Jishi\cite{PRB48} belongs (see
Fig. 2). 
\epsfxsize 6.5 cm
\begin{figure}
\centerline{\epsfbox{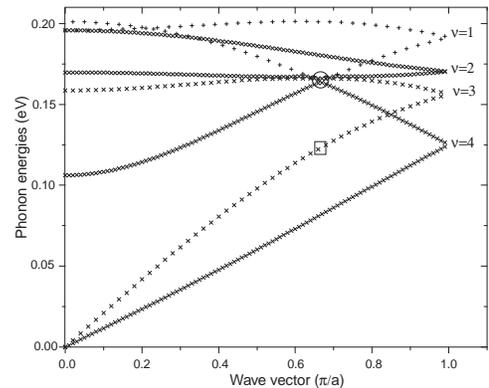}}
\caption{ Phonon dispersion of the (5,5) nanotube coming from the folded-in%
$\vec{q}=(0,q_{y})$ and the $\vec{q}=(2\pi /3^{\frac{1}{2}}a,q_{y})$ phonons of
graphene which coalesce at $q=2\pi /3a$. The circle contains the mode that
couples the electrons at the Fermi level. The square shows the longitudinal
mode previously used by other authors.}
\label{}
It should be apparent by now that the $K$ point phonons naturally
couple to the K\'{e}kul\'{e} structure whereas the acoustic phonons, having
strictly longitudinal polarization, cannot couple to the transverse (around
the nanotube) part of the K\'{e}kul\'{e} modulation. Indeed, the\ circular
polarization (order parameter) associated with the $\left| q_{y}\right|
=\left| q+G_{q}\right| =4\pi /3a$ and $m=0$ phonon, as can be deduced from
the polarizations above, corresponds to a K\'{e}kul\'{e} modulation. To show
this, let us assume a non-zero order parameter $\left\langle
b_{-k_{F},1,0}+b_{k_{F},1,0}^{+}\right\rangle =b_{KM}e^{i\phi }$ giving rise
to a displacement ${\bf u}_{n,s}=\sqrt{\hbar \omega _{k_{F},1,0}/2}\;{\bf S}%
_{s}(-k_{F},1,0)\exp \left[ -i\left( k_{F}R_{n}-\phi \right) \right]
b_{KM}+c.c.$ One obtains the following relative displacements ${\bf d}%
_{n,s,n^{\prime },s^{\prime }}={\bf u}_{n,s}-{\bf u}_{n^{\prime },s^{\prime
}}$: 
\begin{eqnarray}
{\bf d}_{n,3,n,4} =&&-D\;\left( 1,0\right) \cos \left[ \alpha \left(
n\right) \right] , \cr
{\bf d}_{n,2,n,3} =&&D\;\left( \cos \left( \pi /3\right) ,\sin \left( \pi
/3\right) \right) \cos \left[ \alpha \left( n\right) +\pi /3\right] , \cr
{\bf d}_{n,1,n,2} =&&-D\;\left( 1,0\right) \cos \left( \alpha \left(
n\right) +2\pi /3\right) , \cr
{\bf d}_{n-1,3,n,2} =&&D\;\left( -\cos \left( \pi /3\right) ,\sin \left( \pi
/3\right) \right) \cos \left[ \alpha \left( n\right) +\pi \right] ,
\end{eqnarray}
where $D=2\sqrt{\hbar \omega _{-k_{F},1,0}/20}\;b_{KM}$, and $\alpha \left(
n\right) =\left( n-1\right) 2\pi /3+\pi /2-\phi $. This is the
K\'{e}kul\'{e} modulation with period $\Delta n=3$ or wavelength $\Delta
y=3a $. The relative displacements lie along the carbon-carbon bond
directions leading to maximum modulation of the electron hopping amplitude $%
J_{0}$ and thus maximum electron-phonon coupling.

In the situation were $k$ ( $k^{\prime }$) is near $+k_{F}$ ($-k_{F}$) is,
for the $K$ point phonons having $m=0,5$, one finds

\begin{equation}
g_{k,1;k^{\prime },2}^{v,m}(\frac{2\pi }{a})=g_{k,2;k^{\prime },1}^{v,m}(%
\frac{2\pi }{a})=0\ \ \ \ \forall \nu =1,...,4 
\end{equation}
for $p\neq p^{\prime }$ and 
\begin{eqnarray}
g_{k,p;k^{\prime },p}^{3,m}(\frac{2\pi }{a})= && g_{k,p;k^{\prime },p}^{4,m}(%
\frac{2\pi }{a})=0\text{,}\cr 
g_{k,p;k\prime ,p}^{2,m}(\frac{2\pi }{a}) =&&(-1)^{p+1}\frac{4i\eta _{2}}{%
\sqrt{40}} \left[ \sin \left( \frac{k-k^{\prime }}{4}a-\frac{\pi }{6}\right)\right.
\cr
&& \left.\times \cos \left( \frac{k+k^{\prime }}{4}a\right) -\frac{1}{2}\right] 
 \text{ ,}\cr
g_{k,p;k^{\prime },p}^{1,m}(\frac{2\pi }{a}) =&&(-1)^{p}\frac{4i\eta _{1}}{%
\sqrt{40}}\left[ \sin \left( \frac{k-k^{\prime }}{4}a+\frac{\pi }{6}\right)
\right. \cr
&&\left. \times \cos \left( \frac{k+k^{\prime }}{4}a\right) +\frac{1}{2}%
\right] \text{\ ,}
\end{eqnarray}
for $p=p^{\prime }=1,2$. Here $\eta _{\nu }=q_{o}J_{o}\sqrt{\hbar
a/2m_{c}\omega _{0}}$ ($\omega _{0}=\omega _{\pm 2\pi /3a,\nu ,0}$), where $%
m_{c}$ is the mass of a carbon atom, and $J(\vec{r})$ is assumed to depend
only on distance such that $q_{0}J_{0}=\left[ \partial J(\left| \vec{r}%
\right| )/\partial \left| \vec{r}\right| \right] _{0}$\cite{PRL25}. It can
be seen that $g_{k,1;k^{\prime },1}^{2,m}(2\pi /a)=0$ and $g_{k,1;k^{\prime
},1}^{1,m}(2\pi /a)=-6i\eta _{1}/\sqrt{40}$ at $k=-k^{\prime }=k_{F}$.

Futhermore it can be demonstrated that

\begin{equation}
g_{k,1;k^{\prime },1}^{v,0}(\frac{2\pi }{a})=g_{k,2;k^{\prime },2}^{v,0}(%
\frac{2\pi }{a})=0\text{ \ \ \ \ \ \ }\forall \text{ }\nu =1,...,4\text{.}
\end{equation}
for all $m=0$ and $\left| q\right| =2\pi /3a$ phonons (this includes the
acoustic phonon). This result is different from the estimate of the other
mentioned authors. We recover their result when evaluating $g_{k,1;k^{\prime
},1}^{v,0}(0)$ or, equivalently, assuming the longitudinal polarization
corresponds to a $K$ edge phonons. This is probably at the origin of the
confusion.

One can now estimate the mean-field transition temperature. After
integrating the phonons in the interaction representation, the partition
function of the system can be expressed as

\begin{equation}
Z=Z_{ph}^{0}{\rm Tr}\left\{ e^{-\beta H_{0,e}}\text{ }T_{\tau }\text{ }\left[
e^{-\frac{\beta }{2}\int_{0}^{\beta }H_{I}\left( \tau -\tau ^{\prime
}\right) d\left( \tau -\tau ^{\prime }\right) }\right] \right\} 
\end{equation}
where $H_{0,e}=\sum_{p,k,\sigma }\varepsilon _{p}\left( k\right)
C_{k,p,\sigma }^{+}C_{k,p,\sigma }$, and $H_{I}\left( \tau -\tau ^{\prime
}\right) $ is the retarded electron-electron interaction given by%

\begin{eqnarray}
H_{I}\left( \tau -\tau ^{\prime }\right) &&  =\frac{1}{L}\sum_{p,k,\sigma
;p^{\prime },k^{\prime },\sigma ^{\prime };q}
 D_{0}(q,\tau -\tau ^{\prime
}) \cr
&&\times \  g_{k,p;k^{\prime },p}^{1,0}(\frac{2\pi }{a})\left[ g_{k,p^{\prime
};k^{\prime },p^{\prime }}^{1,0}(\frac{2\pi }{a})\right] ^{\ast
}\cr
&&\times \  C_{k,p,\sigma }^{+}C_{k-q-\frac{2\pi }{a},p,\sigma }C_{k^{\prime
},p^{\prime },\sigma ^{\prime }}^{+}C_{k^{\prime }+q+\frac{2\pi }{a}%
,p^{\prime },\sigma ^{\prime }}. 
\end{eqnarray}
This last expression can be alternately written as

\begin{equation}
H_{I}\left( \tau -\tau ^{\prime }\right) =\sum_{p,k,\sigma ;p^{\prime
},k^{\prime },\sigma ^{\prime };q}g(\tau -\tau ^{\prime })O_{KM}(q,\tau
)O_{KM}^{^{+}}(q,\tau ) 
\end{equation}
where
\begin{equation}
g(\tau -\tau ^{\prime })=D_{0}(-\frac{2\pi }{3a},\tau -\tau ^{\prime
})\left| g_{k_{F},1;-k_{F},1}^{1,0}(\frac{2\pi }{a})\right| ^{2}\text{, } 
\end{equation}
\begin{eqnarray}
O_{KM}(q,\tau ) =&&\frac{1}{\sqrt{L}}\sum_{k,\sigma }\left[ C_{1,\sigma
}^{+}\left( k\right) C_{1,\sigma }\left( k-q-\frac{2\pi }{a}\right) \right.
\cr
&&\left. -C_{2,\sigma }^{+}\left( k\right) C_{2,\sigma }\left( k-q-\frac{%
2\pi }{a}\right) \right] \text{ , }
\end{eqnarray}
and the Matsubara time Fourier transform of the phonon propagator is $%
D_{0}(q,\omega _{M})=-2\omega _{q,1,0}/\left( \omega _{q,1,0}^{2}+\omega
_{M}^{2}\right) $. The order parameter associated with $O_{KM}$, the
operator that is naturally\ selected by the electron-phonon coupling, is the
electronic density K\'{e}kul\'{e} modulation operator. To verify this, let
us calculate the electron density $\left\langle \Psi ^{\ast }(\vec{r})\Psi (%
\vec{r})\right\rangle $ where $\Psi (\vec{r})=\sum_{k,p,\sigma
,n,s}e^{iky_{n}}U_{s,p}(k)\phi (\vec{r}-\vec{R}_{n,s})\;C_{k,p,\sigma }$, $%
\phi (\vec{r}-\vec{R}_{n,s})$ is the $\pi $ orbital of the carbon atom at $%
\vec{R}_{n,s}$, and the average is taken over a broken symmetry situation
having $<O_{KM}>=\frac{1}{\sqrt{L}}\sum_{\sigma }\left( \left\langle
C_{k_{F},1,\sigma }^{+}C_{-k_{F},1,\sigma }\right\rangle -\left\langle
C_{k_{F},2,\sigma }^{+}C_{-k_{F},2,\sigma }\right\rangle \right) =\rho
_{KM}e^{i\varphi }$. One can show that 
\begin{eqnarray}
\left\langle \Psi ^{\ast }(\vec{r})\Psi (\vec{r})\right\rangle =&&\frac{1}{2}%
\sum_{n,s\neq n^{\prime },s^{\prime }}\kappa (n,s;n^{\prime },s^{\prime
})\;\rho _{KM}\; \cr
&&\times \phi ^{\ast }(\vec{r}-\vec{R}_{n^{\prime },s^{\prime }})\phi (\vec{r%
}-\vec{R}_{n,s})\;,
\end{eqnarray}
in which only the bond densities contribute $(n,s\neq n^{\prime },s^{\prime
})$. One has $\kappa (n,3;n,4)=\frac{4}{10}\cos \left[ n2\pi /3-\varphi %
\right] ,$ $\kappa (n,2;n,3)=-\frac{4}{10}\cos \left[ \left( n+1/2\right)
2\pi /3-\varphi \right] ,$ $\kappa (n,1;n,2)=\frac{4}{10}\cos \left[ \left(
n+1\right) 2\pi /3-\varphi \right] ,$ and $\kappa (n,2;n-1,3)=-\frac{4}{10}%
\cos \left[ \left( n+3/2\right) 2\pi /3-\varphi \right] .$ Comparing these
bond density modulations to the relative phonon displacements calculated
earlier, one sees that the two coincide when putting $\phi =\varphi +\pi
/2-2\pi /3$.

We can now calculate the\ static K\'{e}kul\'{e} modulation susceptibility
for the system at $q=2\pi /3a$. The random-phase approximation (RPA) result
is given by :

\begin{equation}
\chi_{KM}=\frac{\chi }{1-g(\omega _{M}=0)(\chi _{0}^{1}+\chi _{0}^{2})} 
\end{equation}
where $\chi _{0}^{1}$ ($\chi _{0}^{2}$) is the bare RPA charge response
function of the first (second) band , $\chi _{KM}=\chi ^{1}+\chi ^{2}$ , and 
$\chi =\chi _{0}^{1}+\chi _{0}^{2}-2g(0)\chi _{0}^{1}\chi _{0}^{2}$ . As $%
g(0)$ and $\chi _{0}^{1}+\chi _{0}^{2}$ are both negative, the
susceptibility can diverge when the denominator goes to zero. 
This correspond to a mean-field critical temperature given by

\begin{equation}
T_{c}=3.86\text{ }J_{0}\exp \left( \frac{1}{2N\left( \varepsilon _{F}\right)
g(0)}\right) \text{.} 
\end{equation}

To obtain this pre-exponential constant, we did a numerical calculation of $%
\chi _{0}^{1}$ ($\chi _{0}^{2}$) using the exact dispersion given above and
fitted it to a standard $C\ln (D/T)$ form. Here, one has $N\left(
\varepsilon _{F}\right) =2/\left( \pi \sqrt{3}aJ_{0}\right) $ and $g(0)=-%
\frac{9}{10}$ $\left( q_{o}^{2}\text{ }J_{o}^{2}\text{ }a\right) /\left(
m_{c}\omega _{0}{}^{2}\right) $. For a numerical estimate of $T_{c}$, we
take \cite{Solide} $a=2.46%
\mathop{\rm \AA }%
$, $J_{0}=2.2%
\mathop{\rm eV}%
$, $q_{0}=2.5%
\mathop{\rm \AA }%
^{-1}$, along with our value of $\hbar \omega _{0}=0.166%
\mathop{\rm eV}%
$ corresponding to the graphene $K$ point phonon. We obtain $Tc=15.2$ K.
This estimated Peierls temperature is of the same order as Huang's even
though the selected phonon mode is different. The reason is that although
the numerical factor for our effective electron-phonon coupling amplitude is
larger, our phonon frequency is also larger. The two effects nearly cancel
each other.

Even though nothing much has changed for the numerical estimate of the
Peierls-K\'{e}kul\'{e} mean-field transition, we feel it is important that
people use the correct electron-phonon interaction coupling to the right
phonon. This is important in conductivity or superconductivity calculations.

The predicted $T_{C}$ would of course change if one were to take into
account the modification in stiffness constants due to the curvature of the
SWCNT. These changes would be quantitative, not qualitative. One may of
course, because of the Mermin-Wagner theorem, question the validity of a
mean-field transition in 1D systems. There is of course no transition. The
mean-field $T_{C}$ signals the occurrence of a pseudo-gap\cite{C1}.
Moreover, the mean-field formula is for adiabatic phonons. This is surely
not the case here as the thermal energy at the putative transition $%
k_{B}T_{C}$ is much smaller that the phonon energy. Quantum fluctuation
effects would be considerable. They would wipe out the pseudo-gap.

The above analysis also presumes of an infinitely long nanotube. These are
of course of finite (micron scale) length. The analysis would prevail as
long as the coherence distance is smaller than the tube's length.

\def\btt#1{{\tt$\backslash$#1}}
\def\BibTeX{\rm B{\sc ib}\TeX}

\end{figure}

\end{document}